\documentclass[aps,prx,twocolumn,showpacs,superscriptaddress,longbibliography]{revtex4-2}
\usepackage{amsmath, amsfonts, amssymb,graphicx}
\usepackage{graphicx,epstopdf}
\usepackage{gensymb}
\usepackage[dvipsnames]{xcolor}
\newcommand{\be}{\begin{equation}}
\newcommand{\ee}{\end{equation}}
\newcommand{\bea}{\begin{eqnarray}}
\newcommand{\eea}{\end{eqnarray}}
\newcommand{\bse}{\begin{subequations}}
\newcommand{\ese}{\end{subequations}}

\usepackage{multibib}
\usepackage{color}
\usepackage[colorlinks,bookmarks=false,citecolor=darkblue,linkcolor=red,urlcolor=blue]{hyperref}

\definecolor{darkred}{rgb}{0.7,0.0,0.0}

\definecolor{darkblue}{rgb}{0,0.02,0.45}

\definecolor{darkgreen}{rgb}{0.02,0.45,0.0}

\definecolor{violet}{rgb}{0.8,0.2,0.6}

\begin{document}

\preprint{APS/123-QED}

\title{Spin-orbit-entangled frustrated magnetism in fcc Ba$_2$(Yb,Nd)NbO$_6$ double perovskites}

\author{S. M. Hossain}
\affiliation{Department of Physics, Shiv Nadar Institution of Eminence, Gautam Buddha Nagar, Uttar Pradesh 201314, India}

\author{Sk. Soyeb Ali}
\address{Department of Physics, Bennett University, Greater Noida 201310, Uttar Pradesh, India}
 
\author{S. Mohanty}
\address{School of Physics, Indian Institute of Science Education and Research, Thiruvananthapuram-695551, India}

\author{R. Kolay}
\address{School of Physics, Indian Institute of Science Education and Research, Thiruvananthapuram-695551, India}

\author{M. P. Saravanan}
\affiliation{UGC-DAE Consortium for Scientific Research, University Campus, Khandwa Road, Indore, 452001, India}

\author{A. K. Yogi}
\affiliation{UGC-DAE Consortium for Scientific Research, University Campus, Khandwa Road, Indore, 452001, India}

\author{Y. Tokiwa}
\affiliation{Advanced Science Research Center, Japan Atomic Energy Agency, Tokai, Ibaraki 319-1195, Japan}

\author{R. Nath}
\affiliation{School of Physics, Indian Institute of Science Education and Research, Thiruvananthapuram-695551, India}

\author{S. K. Panda}
\address{Department of Physics, Bennett University, Greater Noida 201310, Uttar Pradesh, India}

\author{M. Majumder}
\email{mayukh.majumder@snu.edu.in}
\affiliation{Department of Physics, Shiv Nadar Institution of Eminence, Gautam Buddha Nagar, Uttar Pradesh 201314, India}

\date{\today}           

\begin{abstract}
The search for candidate Kitaev materials has largely focused on 4$d$ and 5$d$ transition-metal compounds with various lattice geometries. In contrast, investigations of rare-earth 4$f$ systems have thus far been restricted mainly to honeycomb and triangular lattices. In this work, we investigate the rare-earth-based double perovskites Ba$_2$YbNbO$_6$ and Ba$_2$NdNbO$_6$, which crystallize in a face-centered cubic structure. Magnetization and heat-capacity measurements establish isolated ${j_{\rm eff}} = 1/2$ Kramers doublet ground states arising from strong spin-orbit coupling (SOC) and crystal electric-field effects, which are further supported by density-functional theory calculations. Millikelvin-temperature thermodynamic measurements reveal long-range magnetic order with moderate frustration in both compounds. The emergence of magnetic order may be understood within an order-by-disorder scenario, as theoretically proposed for rare-earth fcc lattices with finite Kitaev interactions. Our results thus identify Ba$_2$YbNbO$_6$ and Ba$_2$NdNbO$_6$ as promising rare-earth spin-orbit-entangled magnets and motivate further experimental and theoretical investigations aimed at determining the complete exchange tensor to elucidate the microscopic origin of the underlying magnetic interactions.

\end{abstract}
                            
\maketitle

\section{Introduction}
The ground state wavefunction of a quantum spin liquid (QSL) is inherently nontrivial, characterized not by a single configuration but by a quantum superposition of many degenerate states~\cite{Balents2010}. In the seminal contribution in 2006, Alexei Kitaev introduced an exactly solvable mathematical model of spin-1/2 magnetic ions arranged on a two-dimensional honeycomb lattice, with an exactly solvable QSL ground state~\cite{KITAEV20062}. The framework to realize in real materials with high SOC was initially developed by G. Khaliullin by considering Ising spins located in an edge-sheared octahedral environment of ligands. Instead of conventional geometrical frustration, anisotropic bond-dependent ``exchange frustration" can be realized here. 
This model demonstrated that in edge-sharing octahedra, the direct exchange pathway is suppressed, favouring anisotropic superexchange interactions via the ligand~\cite{10.1143/PTPS.160.155,PhysRevLett.102.017205}. Kitaev spin liquids are particularly intriguing because they possess an exactly solvable ground state with fractionalised excitations in the form of emergent Majorana fermions, quasiparticles with significant interest in the context of fault-tolerant topological quantum computation~\cite{KITAEV20032,PhysRevLett.104.040502,Nayak20081083}. A crucial requirement for realizing the Kitaev interaction is the presence of an effective $j_{\rm eff}=1/2$ spin state, typically stabilized by strong spin-orbit coupling (SOC). As a result, most Kitaev materials have been identified among 4\textit{d} and 5\textit{d} transition metal compounds, where SOC plays a dominant role.

The first experimental realization of a $j_{\rm eff}=1/2$ spin-orbit-entangled Mott insulating state was reported in Sr$_2$IrO$_4$, a compound structurally analogous to the high-\textit{T}$_\mathrm{C}$ cuprate superconducting parent material La$_2$CuO$_4$~\cite{PhysRevLett.101.076402,doi:10.1126/science.1167106}. Despite similarities in magnetic ordering and electronic structure, the underlying mechanisms governing the ground states are different~\cite{doi:10.1126/science.1251151,PhysRevLett.115.176402}. Following this discovery, Kitaev physics has been explored in related Mott insulators such as Na$_2$IrO$_3$~\cite{PhysRevB.85.180403,PhysRevB.83.220403,PhysRevB.82.064412}, $\alpha$-Li$_2$IrO$_3$~\cite{PhysRevLett.108.127203,PhysRevB.93.195158}, $\beta$-Li$_2$IrO$_3$~\cite{PhysRevLett.114.077202,Ruiz2017}, $\gamma$-Li$_2$IrO$_3$~\cite{PhysRevLett.113.197201}, and $\alpha$-RuCl$_3$~\cite{PhysRevLett.114.147201,PhysRevB.93.134423,PhysRevB.90.041112,PhysRevB.92.235119,doi:10.1126/science.aah6015,Bruin2022}, where magnetic ions form well-separated honeycomb planes. More recently, new sets of iridate compounds, H$_3$LiIr$_2$O$_6$~\cite{Kitagawa2018}, Cu$_3$LiIr$_2$O$_6$~\cite{https://doi.org/10.1002/pssb.202100146}, Ag$_3$LiIr$_2$O$_6$~\cite{PhysRevResearch.4.033025}, and Li$_2$RhO$_3$~\cite{Shen2025}, with layered honeycomb structures, expand the family of honeycomb-Kitaev materials. Interestingly, Na$_3$Co$_2$SbO$_6$, containing Co$^{2+}$ ions in the high-spin state, despite lacking the original Jackeli-Khaliullin conditions, has also exhibited signatures of dominant Kitaev interactions~\cite{PhysRevB.97.014407,PhysRevB.102.224429,PhysRevB.97.014407}. These observations have prompted a reframing of Kitaev magnetism beyond the Jackeli-Khaliullin formulation by H. Liu and G. Khaliullin~\cite{PhysRevB.97.014407}. Notably, rare-earth (4\textit{f}) systems have attracted growing interest in this regard~\cite{PhysRevB.99.241106}. The strongly localized nature of 4\textit{f} orbitals enhances single-ion anisotropy and naturally favors Ising-type behavior, enabling bond-directional interactions, not commonly accessible in 3\textit{d} or 4\textit{d} ions. In contrast to extended \textit{d} orbitals, which often promote long-range ordering via further-neighbor interactions, 4\textit{f}-based materials may better support the realization of a true Kitaev spin liquid. While Kitaev interactions have been extensively investigated in \textit{d}-block elements, the \textit{f}-block systems with different lattice geometries remain largely unexplored. Only, a very few frustrated materials based on \textit{f}-block elements such as, Na$_2$PrO$_3$ (honeycomb)~\cite{PhysRevB.110.064425}, CsCeSe$_2$ (triangular)~\cite{PhysRevLett.133.096703}, and YbOCl (triangular)~\cite{PhysRevResearch.6.033274} are explored till date.

A recent theoretical proposal has predicted the emergence of dominant Kitaev interactions in rare-earth double perovskite compounds where magnetic ions form a face-centered cubic (fcc) lattice. In these systems, due to the presence of finite Kitaev interactions, exotic Weyl-magnon excitation has been predicted below the ordering temperature~\cite{PhysRevB.95.085132}. Interestingly, a QSL state can only be achieved in the absence of Kitaev exchange with finite nearest-neighbor and next-nearest-neighbor~\cite{PhysRevB.100.085139}. However, so far, no detailed exploration of magnetic properties in this direction has been reported. In this context, Ba$_2$YbNbO$_6$ and Ba$_2$NdNbO$_6$ stand out as two promising candidates for exploring spin-orbit-entangled magnetism in rare-earth-based double perovskites. Here, Yb$^{3+}$ and Nd$^{3+}$ ions are in an octahedral environment that forms a 3D fcc sublattice of edge-sharing triangles, a configuration that effectively suppresses direct exchange and enhances bond-dependent interactions. The presence of strong crystal electric field (CEF) effects and SOC, combined with geometric frustration inherent to the lattice, provides an ideal stage for the emergence of Kitaev-like physics in the ground state Kramer's doublet with $j_{\rm eff}=1/2$.

\section{METHODOLOGY}
Polycrystalline insulating compounds Ba$_2$YbNbO$_6$ and Ba$_2$NdNbO$_6$ were synthesized using the conventional solid-state reaction method~\cite{Maity_2013}. Initially, Yb$_2$O$_3$ and Nd$_2$O$_3$ were dried overnight at 800$\degree$C to remove moisture content. An appropriate stoichiometric amount of BaCO$_3$ (Sigma-Aldrich, 99.99$\%$), Nb$_2$O$_5$ (Sigma-Aldrich, 99.995$\%$), Yb$_2$O$_3$ (Sigma-Aldrich, 99.99$\%$), and Nd$_2$O$_3$ (Sigma-Aldrich, 99.99$\%$), is homogeneously mixed separately for two compounds using a high-purity ethanol solution and ground for several hours. The stoichiometric mixtures were then placed in a high-purity alumina crucible and heated in air at 900$\degree$C for 12 hours for degassing. Finally, the mixtures were compressed into pellets and sintered at 1300$\degree$C for 24 hours with several intermediate grindings. At each step, powder X-ray diffraction (PXRD) was performed at room temperature using a Bruker D8 Advanced powder diffractometer with monochromatic CuK$_\alpha$ radiation ($\lambda_{ave}=1.54056$~\AA, 40 kV/40 mA) to monitor the systematic evolution of the desired phase. The absence of any extra peaks indicates phase purity of the samples.

\begin{figure}
      \centering
      \includegraphics[width=0.48\textwidth]{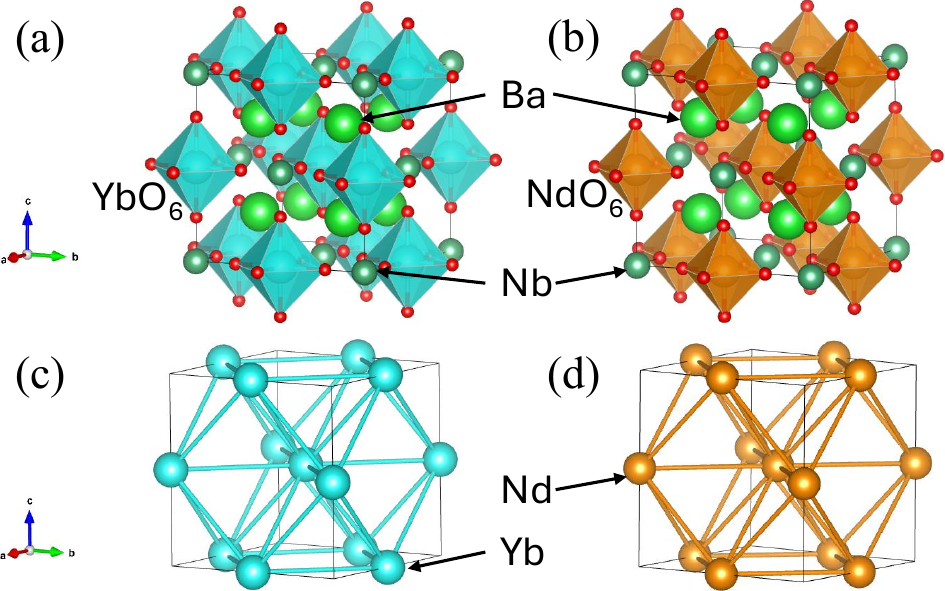}
      \caption{(a)-(b) Crystal structure of Ba$_2$YbNbO$_6$ and Ba$_2$NdNbO$_6$ with cubic symmetry space group \textit{Fm}-3\textit{m}. The cyan and golden coloured octahedron formed by the \textit{RE} ions, Yb and Nd, respectively, with six O atoms in red dots. (c)-(d) The arrangement of magnetic \textit{RE} ions, Yb and Nd, in a fcc lattice geometry.}
      \label{fig: Crystal structure}
\end{figure} 

Temperature-dependent DC magnetization in the range of 0.4~K to 400~K at different applied magnetic fields was measured using a superconducting quantum interference device (SQUID) magnetometer (MPMS-3, Quantum Design). Isothermal magnetization curves were obtained at different constant temperatures (\textit{T} = 0.4~K, 0.6~K, 0.8~K, 1.8~K, 3~K, and 5~K) with varying magnetic fields from 0~T to 7~T. Below 1.8~K, measurements were performed using a ${}^{3}\mathrm{He}$ (iHelium3) insert to the SQUID magnetometer. Heat capacity in the temperature range 0.4~K$\leq T \leq$300~K at different constant applied magnetic fields of 0-9~T on high-temperature sintered pellets was measured using the thermal relaxation technique in a Physical Property Measurement System (PPMS). Below 2~K, measurement was performed using a ${}^{3}\mathrm{He}$ (iHelium3) insert to PPMS. Furthermore, the zero-field heat capacity down to 0.1~K was measured in a dilution refrigerator.

\begin{figure}
      \centering
      \includegraphics[width=0.48\textwidth]{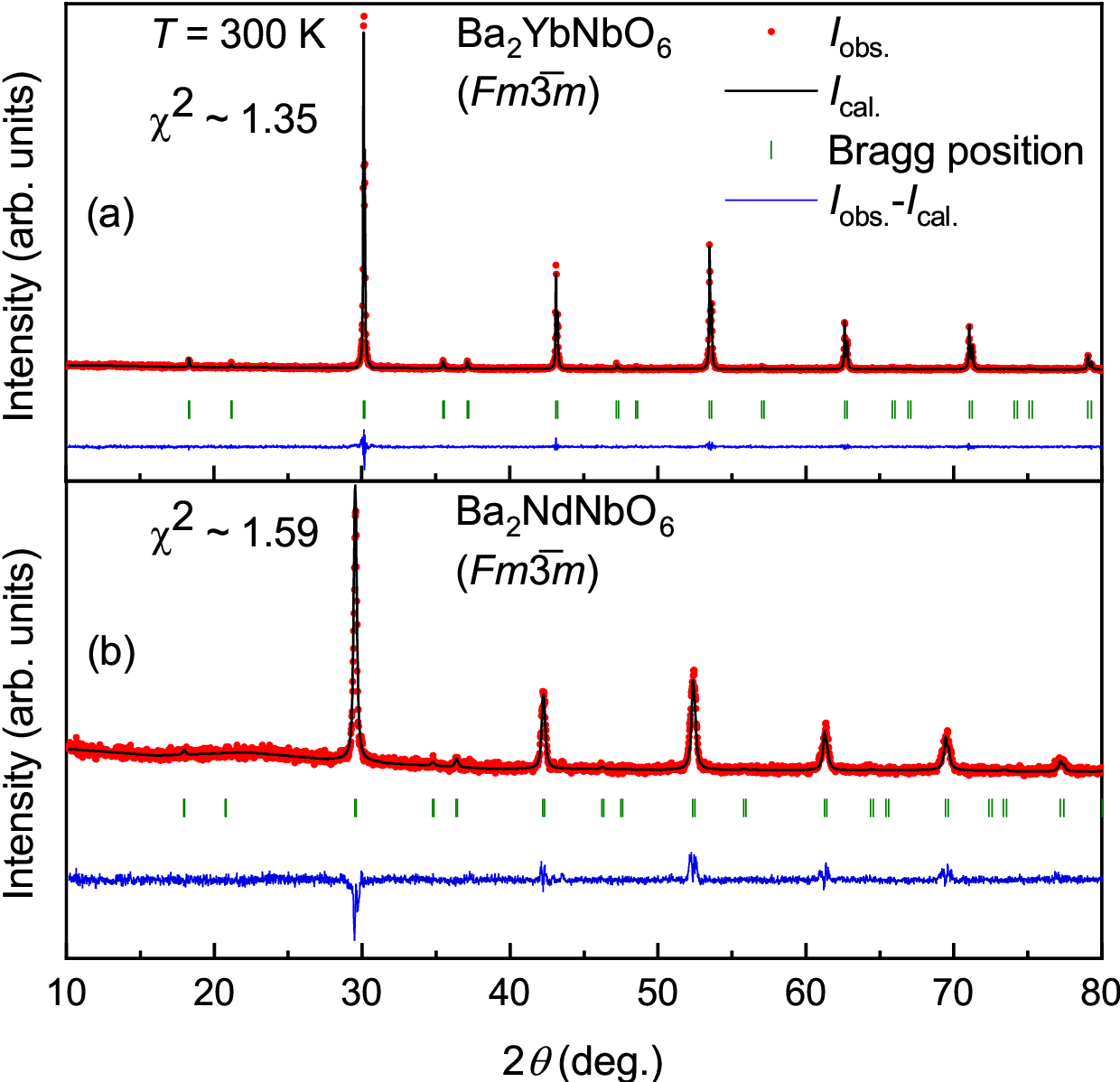}
      \caption{(a)-(b) Rietveld refinement of the room-temperature PXRD patterns of Ba$_2$YbNbO$_6$ and Ba$_2$NdNbO$_6$, respectively. Red dots represent the observed intensities, and the black solid line shows the calculated PXRD pattern. The green bar symbols represent the Bragg peak positions corresponding to the calculated pattern. A difference between the Rietveld fit of the calculated PXRD profile and the observed intensity is shown by the blue solid line. $\chi^2$ value indicates the quality of goodness-of-fit of the Rietveld refinement.}
      \label{fig: PXRD}
\end{figure} 

Density functional theory (DFT)~\cite{DFT1, DFT2} calculations were carried out using the full-potential linearized augmented plane-wave (FP-LAPW) method as implemented in the \textsc{WIEN2K}~\cite{wien2k} code. The exchange-correlation energy was treated within the generalized gradient approximation (GGA) using the Perdew-Burke-Ernzerhof (PBE)~\cite{GGA_PBE} functional. To account for the strong electronic correlations associated with the localized $4f$ orbitals of the rare-earth ions (Yb and Nd), we employed the GGA+$U$~\cite{GGA+U} approach. The on-site Coulomb interaction and Hund’s exchange parameters were chosen as $U = 6$~eV and $J_H = 0.5$~eV, respectively, following values widely used in earlier theoretical studies on similar rare-earth systems~\cite{Hubbard-U}. These parameters ensure a realistic description of the correlated $4f$ manifold while avoiding both over-localization and spurious delocalization. Within the FP-LAPW framework, the unit cell is divided into non-overlapping muffin-tin (MT) spheres centered on each atom and an interstitial region outside the spheres. Inside the MT spheres, the Kohn-Sham wave functions were expanded in spherical harmonics up to an angular-momentum cutoff of $l_{\mathrm{max}} = 10$, ensuring an accurate representation of the atomic-like states. In the interstitial region, plane waves were used with appropriately converged basis-set parameters. Since the compounds under investigation contain heavy elements for which relativistic effects can significantly influence the low-energy electronic structure, spin-orbit coupling (SOC) was incorporated using the second-variational scheme implemented in \textsc{WIEN2K}. The combined treatment of electronic correlations and relativistic effects will henceforth be referred to as the GGA+$U$+SOC approach. Next, to quantify the magnetic exchange interactions, we performed GGA+$U$+SOC calculations using the full-potential linear muffin-tin orbital (FP-LMTO) method as implemented in the \textsc{RSP}t code. After achieving self-consistency, the magnetic force theorem was employed to extract the effective inter-site exchange coupling parameters ($J_{ij}$)
 between the Yb ions. Details of the $J_{ij}$ calculations using FP-LMTO implementation in \textsc{RSP}t can be found in Refs.~\cite{Rspt1,Rspt2,Rspt3}. This approach provides an efficient and reliable framework for evaluating magnetic exchange interactions and has been successfully applied to a wide range of transition-metal and rare-earth compounds~\cite{PhysRevB.111.195148,x8cf-x6cv}.

\section{RESULTS AND DISCUSSIONS}

\subsection{Crystal Structure}
The desired compounds, Ba$_2$YbNbO$_6$ and Ba$_2$NdNbO$_6$, are crystallized in cubic symmetry with space group \textit{Fm}-3\textit{m} (space group no. 225) [Fig.~\ref{fig: Crystal structure}(a,b)]. The structural analysis is performed by the Rietveld refinement method using the FULLPROF SUITE software package~\cite{RODRIGUEZCARVAJAL199355}. The obtained lattice parameters are $a=b=c=8.3839(2)$~Å and $a=b=c=8.5546(6)$~Å with $\alpha=\beta=\gamma=90^{\circ}$ for Ba$_2$YbNbO$_6$ and Ba$_2$NdNbO$_6$, respectively. The distance between any two Yb$^{3+}$ and Nd$^{3+}$ ions is $5.928$~\AA~and $6.05$~\AA, respectively, and they make a three-dimensional frustrated fcc lattice arrangement as shown in Fig.~\ref{fig: Crystal structure}(c,d). To assess phase purity, Rietveld refinements have been performed for both structures. Figure~\ref{fig: PXRD}(a,b) shows the full refinement profile and goodness factors performed at room temperature. The results of the Rietveld refinement of the atomic coordinates are summarized in Table~\ref{Yb_refine} and Table~\ref{Nd_refine}. To be noted, the structural refinement using monoclinic and cubic symmetry was rendered to very close reliable weight factors. While the monoclinic model does not provide an improved fit to our room-temperature PXRD data, we acknowledge that neutron diffraction, where the neutron cross-section is larger for lighter elements, may reveal a reduction in symmetry, as observed for Ba$_2$NdNbO$_6$~\cite{SAINES2006187,HENMI1999353}. Although the Rietveld refinement with full occupancy (Occ. = 1.0) for all atomic sites (see Table~\ref{Yb_refine} and Table~\ref{Nd_refine}) is fully consistent with complete $B(B^\prime)$-site rock salt ordering of Yb$^{3+}$/Nd$^{3+}$ and Nb$^{5+}$ on the 4a and 4b Wyckoff positions, respectively~\cite{VASALA20151}. Moreover, our GGA+$U$ calculation indicates that the fcc cubic phase is energetically more favorable, with a total energy of approximately 258 meV/f.u. lower than the monoclinic phase.

\begin{table}[h!]
\centering
\caption{The atomic coordinates of Ba$_2$YbNbO$_6$ were determined through Rietveld refinement of X-ray diffraction data at 300~K considering cubic space group \textit{Fm}-3\textit{m} ($\textit{a=b=c}$ = 8.3846(4)~\AA~and $\alpha=\beta=\gamma$ = 90$\degree$)}
\label{Yb_refine}
\begin{ruledtabular}
\begin{tabular}{cccccc}
Atom & Wyckoff & $x$ & $y$ & $z$ & Occ.\\
     & position & & & &   \\   
\hline 
Ba & 8c & 0.2500 & 0.2500 & 0.2500 & 1 \\
Nb & 4b & 0.0000 & 0.0000 & 0.5000 & 1 \\
Yb & 4a & 0.5000 & 0.0000 & 0.0000 & 1 \\
O & 24e & 0.0000 & 0.0000 & 0.2378(18) & 1 \\
\end{tabular}\end{ruledtabular}
\end{table}

\begin{table}[h!]
\centering
\caption{The atomic coordinates of Ba$_2$NdNbO$_6$ were determined through Rietveld refinement of X-ray diffraction data at 300~K considering cubic space group \textit{Fm}-3\textit{m} ($\textit{a=b=c}$ = 8.5550(7)~\AA and $\alpha=\beta=\gamma$ = 90$\degree$)}
\label{Nd_refine}
\begin{ruledtabular}
\begin{tabular}{cccccc}
Atom & Wyckoff & $x$ & $y$ & $z$ & Occ. \\
      & position & & & &  \\
\hline 
Ba & 8c & 0.2500 & 0.2500 & 0.2500 & 1 \\
Nb & 4b & 0.0000 & 0.0000 & 0.5000 & 1 \\
Nd & 4a & 0.5000 & 0.0000 & 0.0000 & 1 \\
O & 24e & 0.0000 & 0.0000 & 0.2376(59) & 1 \\
\end{tabular}\end{ruledtabular}
\end{table}

The double perovskite structure ($A_2BB^{\prime}$O$_6$) allows octahedral environment for both \textit{B} and $B^{\prime}$ sites. Here, the rare-earth ions (Yb$^{3+}$ and Nd$^{3+}$) are situated in an octahedral environment with parallel edge-shearing geometry, which allows dominant Kitaev interactions as theoretically proposed~\cite{PhysRevB.95.085132}.

\begin{figure*}
      \centering
      \includegraphics[width=0.98\textwidth]{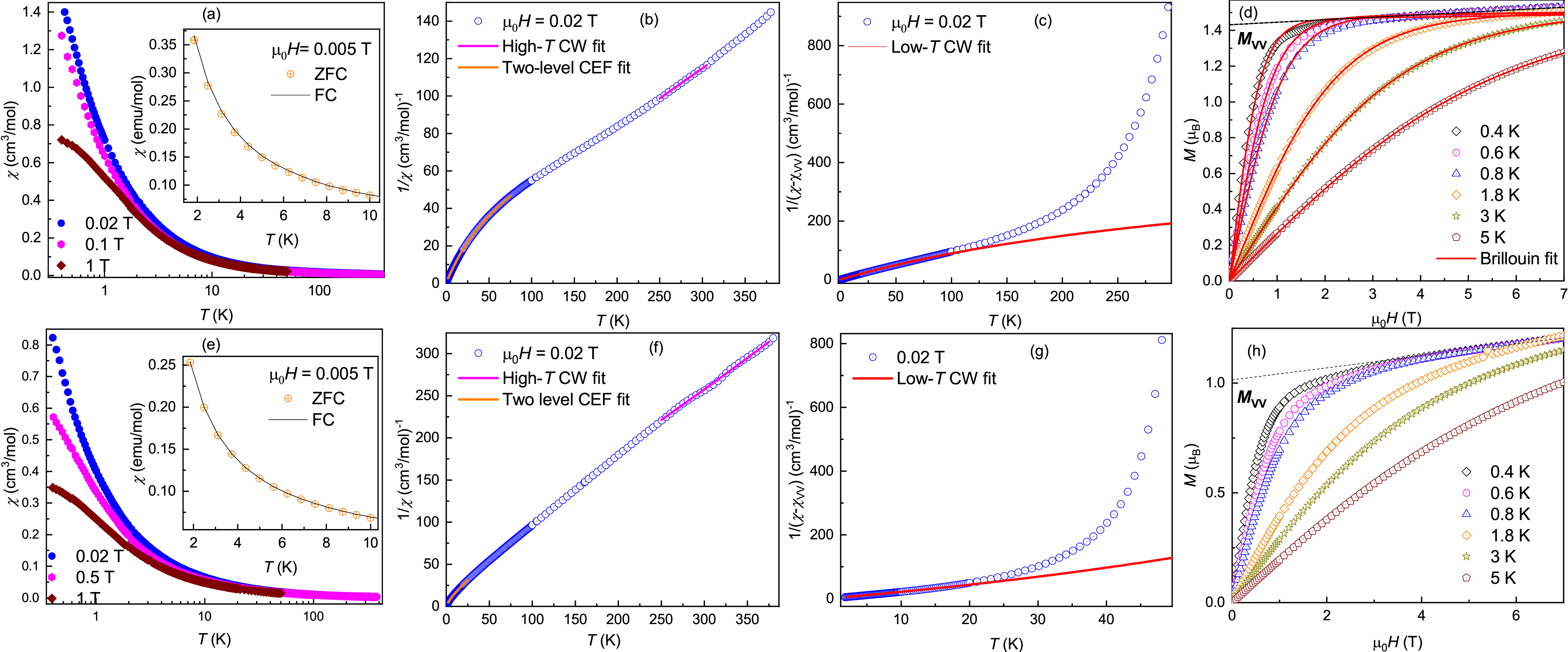}
      \caption{(a) and (e): Temperature-dependent magnetic susceptibility ($\chi \equiv M/H$) from 0.4~K to 400~K measured at different applied magnetic fields (0.02~T, 0.1~T, and 1~T) for Ba$_2$YbNbO$_6$ and (0.02~T, 0.5~T, and 1~T) for Ba$_2$NdNbO$_6$, respectively. Insets showing the absence of spin frozen like transitions 
(b) and (f). Inverse susceptibility data fitted with the Curie-Weiss law in the high temperature limit, and the low temperature fit shows the two-level CEF fit. (c) and (g): The low temperature Curie-Weiss fit to [1/($\chi-\chi_{VV}$)], for Ba$_2$YbNbO$_6$ and Ba$_2$NdNbO$_6$, respectively. (d) and (f): Isothermal magnetization (\textit{M} vs \textit{H}) at various temperatures (0.4~K to 5~K), exhibiting no hysteresis and a tendency toward saturation at high fields for Ba$_2$YbNbO$_6$, and no complete saturation for Ba$_2$NdNbO$_6$.}
      \label{fig: BYNO_magnetization}
\end{figure*} 

\subsection{Magnetization}
Figures~\ref{fig: BYNO_magnetization}(a) and (e) show the temperature dependence of magnetic susceptibility ($\chi \equiv M/H$) down to 0.4~K with different constant applied magnetic fields for Ba$_2$YbNbO$_6$ and Ba$_2$NdNbO$_6$, respectively. With the absence of bifurcation between $\chi(T)$ measured under zero-field-cooled (ZFC) and field-cooled (FC) conditions [insets of Figs.~\ref{fig: BYNO_magnetization}(a) and (e)] at the lowest applied magnetic field of 0.005~T, the possibility of a frozen state is ruled out. No signature of any magnetic long-range ordering (LRO) down to the lowest temperature, 0.4~K, was observed. To obtain a quantitative estimate of the strength of magnetic exchange, the Cuire-Wiess (CW) equation, $\chi(T)=\frac{C}{T-\theta_{CW}}+\chi_0$, has been used to fit the data, as shown in Figs.~\ref{fig: BYNO_magnetization}(b) and (f), indicated by the dashed magenta line for both the compounds Ba$_2$YbNbO$_6$ (250~K$\leq T \leq$305~K) and Ba$_2$NdNbO$_6$ (250~K$\leq T \leq$375~K). In the high temperature region, with an applied magnetic field of 0.02~T, the Curie-Weiss fit renders $\chi_0$, $\mathrm{\textit{C}^{HT}}$, and $\theta_{\mathrm{CW}}^{\mathrm{HT}}$. The value of the temperature-independent term is found to be $\chi_0 \simeq 1.26\times10^{-3}~\mathrm{emu/mol}$ for Ba$_2$YbNbO$_6$ and $\chi_0 \simeq -4.15\times10^{-4}~\mathrm{emu/mol}$ for Ba$_2$NdNbO$_6$, which originates from the contribution of temperature-independent core diamagnetic ($\chi_{\rm dia}$) and/or Van-Vleck paramagnetic ($\chi_{\rm VV}$) susceptibilities. The other two fitting parameters $\mathrm{\textit{C}^{HT}}$ and $\theta_{\mathrm{CW}}^{\mathrm{HT}}$ are the Curie constant and CW temperature, respectively. The value of Curie constant in the high temperature limit $\mathrm{\textit{C}^{HT}}$ $\simeq$ 2.32~emu-K/mol results in an effective moment $\left(\mu_{eff}=\sqrt{8\mathrm{\textit{C}^{HT}}}\right)$ $\sim$ 4.31~$\mu_B$/Yb$^{3+}$, which is close to 4.53~$\mu_B$ for a free Yb$^{3+}$ ion in \textit{J} = 7/2 state also seen in earlier report~\cite{Maity_2013,HENMI1999353}. Similarly for Ba$_2$NdNbO$_6$, the obtained $\mathrm{\textit{C}^{HT}}$ results in $\mu_{eff}\simeq$ 3.64~$\mu_B$/Nd$^{3+}$, which is close to the theoretically calculated value 3.6~$\mu_B$ for free Nd$^{3+}$ ion in \textit{J} = 9/2 state. Table~\ref{MT_analysis} summarizes the resultant parameters. The high temperature  $\theta_{\mathrm{CW}}^{\mathrm{HT}}$ values are -11.38~$\mathrm{K}$ and -87.53~$\mathrm{K}$ for Ba$_2$YbNbO$_6$ and Ba$_2$NdNbO$_6$, respectively, obtained from the fitting. It is important to note that the obtained parameters vary significantly with the fitting temperature window. The magnitude and sign of $\theta_{\mathrm{CW}}^{\mathrm{HT}}$ at high temperatures don't reflect the nature of correlation; it is due to the CEF excitations from ground to higher excited states.


\begin{table}[t]
\centering
\caption{Result of the Curie-Weiss fit to the $\chi(T)$ data at high- and low-temperature regions for Ba$_2$(Yb,Nd)NbO$_6$.}
\vspace{0.1cm}

\setlength{\tabcolsep}{6pt}
\renewcommand{\arraystretch}{1.6}
\begin{tabular}{l c c c c}
\hline \hline
Compound &
\multicolumn{2}{c}{$\theta_{\mathrm{CW}}$ (K)} &
\multicolumn{2}{c}{$\mu_{\mathrm{eff}}$ ($\mu_B$)} \\
\cline{2-5}
 & High $T$ & Low $T$ & High $T$ & Low $T$ \\
\hline
Ba$_2$YbNbO$_6$ & $-11.38$ & $-0.33$ & $4.31$ & $2.69$ \\
Ba$_2$NdNbO$_6$ & $-87.53$ & $-0.52$ & $3.64$ & $2.05$ \\
\hline \hline
\end{tabular}
\label{MT_analysis}
\end{table}

With reducing temperature, $1/\chi$ deviates from linearity, indicating the CEF effect of rare-earth ions (Yb and Nd) situated in an octahedral environment formed by O$^{2-}$ ligands. Being Kramers ions (Yb$^{3+}$ and Nd$^{3+}$), eight-fold (2\textit{J}+1; \textit{J}=7/2) and ten-fold (\textit{J}=9/2) degeneracy of \textit{f}-orbital will be lifted and at least doubly degenerated Kramers pairs formed. At reduced temperatures, higher-energy Kramers levels depopulate, and the lowest Kramers pairs dominate the low-temperature property.  It is important to note that at high temperatures, the Van-Vleck contribution from higher-lying levels is significantly small. Therefore, in low-\textit{T}s, $1/(\chi-\chi_{VV})$ was fitted by the CW equation [shown in Figs.~\ref{fig: BYNO_magnetization}(c) and (g)] for both the cases with $\chi_{\rm VV}$ estimated from the $M$ vs $H$ curves (as discussed later) for Ba$_2$YbNbO$_6$ (4~K $\leq T\leq$ 16~K) and Ba$_2$NdNbO$_6$ (2~K $\leq T \leq$ 10~K). For Ba$_2$YbNbO$_6$, the fitting yields $\mathrm{\textit{C}^{\mathrm{LT}}}\sim 0.904$~emu-K/mol and  $\theta_{\mathrm{CW}}^{\mathrm{LT}}=-0.33~\mathrm{K}$. The negative $\theta_{\mathrm{CW}}^{\mathrm{LT}}$ indicates antiferromagnetic correlations with an effective moment $\mu_{eff}^{\mathrm{LT}}\approx2.68~\mu_B$, which closely matches with other reports~\cite{PhysRevB.108.054442}. This further attributes the oxidation state of Yb to be 3+ with CEF split {${j_{\rm eff}} = 1/2$} ground state Kramers doublet at low-temperature~\cite{Mohanty134408, Somesh064421}. Whereas for Ba$_2$NdNbO$_6$, the fitting results in $\theta_{\mathrm{CW}}^{\mathrm{LT}}\simeq$ -0.52~K with effective moments $\mu_{eff}^{\mathrm{LT}}\simeq$ 2.05~$\mu_B$. The reduced value of $\mu_{eff}^{\mathrm{LT}}$ confirms an effective spin $j_{\rm eff}=1/2$ ground state and matches closely with other results~\cite{PhysRevB.110.144434}. 

To roughly estimate the energy gap $\Delta/{k_{\rm B}}$ between the ground state and the first excited Kramers levels, we have adapted the following equation~\cite{Mugiraneza2022,ydpn-8wgf}:

\begin{equation}
\label{Two_level_magnetization}
\chi(T) = \chi_0 + \frac{1}{8\,(T - \theta_{\mathrm{CW}})}
\left[
\frac{\mu_{\mathrm{eff},0}^2 + \mu_{\mathrm{eff},1}^2
\, e^{-\frac{\Delta}{k_B T}}}
{1 + e^{-\frac{\Delta}{k_B T}}}
\right].
\end{equation}
Here, $\mu_{eff, 0}$ and $\mu_{eff, 1}$ are the effective moments of the ground and first excited states, respectively. The fit in the low-temperature region (2~K$\leq T \leq$75~K) for Ba$_2$YbNbO$_6$ yields $\chi_0 \simeq 0.0085$~emu/mol,  $\mu_{eff, 0} \simeq 2.71$~$\mu_\mathrm{B}$/Yb$^{3+}$, $\mu_{eff, 1} \simeq 2.96$~$\mu_\mathrm{B}$/Yb$^{3+}$, $\Delta/{k_{B}} \simeq 55$~K, and $\theta_{\mathrm{CW}} \simeq - 0.38$~K. Whereas, for Ba$_2$NdNbO$_6$ (2~K $\leq T \leq$25~K) the parameters are $\chi_0 \simeq 0.0044$~emu/mol,  $\mu_{eff, 0} \simeq 2.12$~$\mu_\mathrm{B}$/Nd$^{3+}$, $\mu_{eff, 1} \simeq 2.75$~$\mu_\mathrm{B}$/Nd$^{3+}$, $\Delta/{k_{B}} \simeq 14.4$~K, and $\theta_{\mathrm{CW}} \simeq - 0.55$~K. The resultant parameters closely match those obtained from magnetization and heat capacity, as discussed later sections and in earlier findings~\cite{Maity_2013, HENMI1999353}.

Isothermal magnetization ($M$ vs $H$) curves at $T =$ 0.4~K, 0.6~K, 0.8~K, 1.8~K, 3~K, and 5~K were measured for Ba$_2$YbNbO$_6$ and Ba$_2$NdNbO$_6$ as shown in Figs.~\ref{fig: BYNO_magnetization}(d) and (h), respectively, indicating the paramagnetic nature. A near trend of saturation of isothermal magnetization (\textit{M}) at 0.4~K, for $\mu_0H > 1.5$~T, is observed for Ba$_2$YbNbO$_6$. The slight deviation from saturation behavior is due to the Van Vleck contribution arising from degenerate ground-state Kramers levels, indicating well-isolated ground and first excited states for Ba$_2$YbNbO$_6$. The high-field region of the \textit{M} vs \textit{H} data was fitted by a straight line that results in $\chi_{\rm VV} \simeq 0.01396~\mathrm{\mu_B/T}$. By extrapolating the fit to zero field, an intercept with the $y$-axis gives the saturation magnetization $(\mathrm{\textit{M}_{S}}) \simeq 1.43~\mu_{B}$ corresponding to a \textit{g}-value $\sim$~2.86. This observed \textit{g}-value is slightly enhanced compared to the calculated \textit{g}-value (2.667) of $\Gamma_6$ wavefunction (discuss later), indicating modest CEF-induced mixing while retaining an isotropic Heisenberg-type [Table~\ref{tensor:g}(a)] ${j_{\rm eff}} = 1/2$ ground state. For Ba$_2$NdNbO$_6$ in Fig.~\ref{fig: BYNO_magnetization}(h), no saturation trend in $M$ vs $H$ is observed up to a field of 7~T, which is due to the combined Van Vlack and contribution from the low-laying first CEF level, discussed in the heat capacity result. The Van-Vleck contribution at the high field region is estimated to be  $\chi_{VV} \simeq 2.7\times10^{-6}~\mathrm{\mu_B/T}$ with $\mathrm{\textit{M}_{S}} \simeq 1.01~\mu_{B}$ corresponding to the \textit{g}-value $\sim$~2.02. This further confirms the ${j_{\rm eff}} = 1/2$ ground state with strong magnetocrystalline anisotropy as seen from \textit{DFT} results (discussed below) and also from other studies~\cite{Arh2022}. The anisotropic \textit{g}-tensor [Table~\ref{tensor:g}(b)] indicates the CEF ground state for Nd$^{3+}$ is an easy-plane strong anisotropic XY-type interaction. 

Moreover, the absence of hysteresis in $M$ vs $H$ suggests the absence of ferromagnetic correlations, consistent with the $\chi$ vs $T$ results. In Figs.~\ref{fig: BYNO_magnetization}(d) and (h), a modified Brillouin function: 
\begin{equation}
\label{Brillouin}
M (H) = N_{A}g\mu_BJ_{eff}B_{J_{eff}}(x) + \chi_{VV}H,
\end{equation}
was employed to fit the $M$ vs $H$ isothermal curves.
Here, $B_{J_{eff}}(x)$ is the Brillouin function which can be written as $B_J(x) = \frac{2J + 1}{2J} \coth \left( \frac{2J + 1}{2J} x \right) - \frac{1}{2J} \coth \left( \frac{x}{2J} \right)$ and $x=g\mu_BJ_{\rm eff}H/k_{\rm B}T$. Equation~\eqref{Brillouin} fits well for Ba$_2$YbNbO$_6$ but fails for Ba$_2$NdNbO$_6$. We have fixed $\mathrm{\textit{J}_{eff}=1/2}$ and the fitting yields $\chi_{VV} \simeq 0.0125~\mathrm{\mu_B/T}$ with \textit{g}-value $\sim 2.2$. This low-temperature paramagnetic behavior with \textit{g}-value indicates the single-ion property of Yb$^{3+}$ ions, and the parameters are well-matched with earlier estimations.

\begin{table*}
\caption{Eigenvectors and Eigenvalues of Ba$_2$NdNbO$_6$}
\label{Nd_eigen}
\begin{ruledtabular}
\begin{tabular}{ccccccccccc}
E (meV) &$| -\frac{9}{2}\rangle$ & $| -\frac{7}{2}\rangle$ & $| -\frac{5}{2}\rangle$ & $| -\frac{3}{2}\rangle$ & $| -\frac{1}{2}\rangle$ & $| \frac{1}{2}\rangle$ & $| \frac{3}{2}\rangle$ & $| \frac{5}{2}\rangle$ & $| \frac{7}{2}\rangle$ & $| \frac{9}{2}\rangle$ \tabularnewline
 \hline 
0.000 & 0.0 & 0.0 & 0.5506 & 0.0 & 0.0 & 0.0 & 0.8348 & 0.0 & 0.0 & 0.0 \tabularnewline
0.000 & 0.0 & 0.0 & 0.0 & -0.8348 & 0.0 & 0.0 & 0.0 & -0.5506 & 0.0 & 0.0 \tabularnewline
0.000 & 0.0 & 0.1548 & 0.0 & 0.0 & 0.0 & 0.5925 & 0.0 & 0.0 & 0.0 & -0.7906 \tabularnewline
0.000 & 0.7906 & 0.0 & 0.0 & 0.0 & -0.5925 & 0.0 & 0.0 & 0.0 & -0.1548 & 0.0 \tabularnewline
1.324 & 0.0 & 0.2041 & 0.0 & 0.0 & 0.0 & 0.7638 & 0.0 & 0.0 & 0.0 & 0.6124 \tabularnewline
1.324 & -0.6124 & 0.0 & 0.0 & 0.0 & -0.7638 & 0.0 & 0.0 & 0.0 & -0.2041 & 0.0 \tabularnewline
63.559 & 0.0 & -0.9666 & 0.0 & 0.0 & 0.0 & 0.2562 & 0.0 & 0.0 & 0.0 & 0.0027 \tabularnewline
63.559 & 0.0027 & 0.0 & 0.0 & 0.0 & 0.2562 & 0.0 & 0.0 & 0.0 & -0.9666 & 0.0 \tabularnewline
63.559 & 0.0 & 0.0 & 0.0 & 0.5506 & 0.0 & 0.0 & 0.0 & -0.8348 & 0.0 & 0.0 \tabularnewline
63.559 & 0.0 & 0.0 & -0.8348 & 0.0 & 0.0 & 0.0 & 0.5506 & 0.0 & 0.0 & 0.0 \tabularnewline
\end{tabular}\end{ruledtabular}
\end{table*}

The CEF Hamiltonian gives Kramers and non-Kramers doublets depending on the electronic configuration of the rare-earth ion~\cite{HUTCHINGS1964227}  and can be written as \begin{equation}
\label{CEF:Hamiltonian}
H_{\text{CEF}}=\sum_{i, j} B_{i}^{j} O_{i}^{j}.
\end{equation}
Here, $O_{i}^{j}$ are Stevens operators and $B_{i}^{j}$ are the corresponding CEF parameters to approximate the Coulomb potential generated by the neighboring ligands~\cite{Stevens1952}. By analyzing the point group symmetry of the rare-earth ion, one can make a reasonable estimate of these parameters [Table~\ref{CEF_parameters}]. The single-ion property of Yb$^{3+}$ and Nd$^{3+}$ has been estimated using \texttt{PyCrystalField}~\cite{Scheie:in5044}.

\begin{table}[htbp]
\centering
\caption{Crystal electric field parameters for Ba$_2$YbNbO$_6$ and Ba$_2$NdNbO$_6$.}
\label{CEF_parameters}
\begin{tabular}{ccc}
\hline
Steven's parameter & Ba$_2$YbNbO$_6$ & Ba$_2$NdNbO$_6$ \\
\hline
$B_4^0$ & $-0.04380984$ & $-0.01585894$ \\
$B_4^4$ & $-0.21904921$  & $-0.0792947$ \\
$B_6^0$ & $1.0101\times10^{-4}$ & $-8.021\times10^{-5}$ \\
$B_6^4$ & $-0.00212131$ & $0.00168439$ \\
\hline\hline
\end{tabular}
\end{table}

Since both Yb$^{3+}$ and Nd$^{3+}$ are in an octahedral environment (\textit{O}$_\mathrm{h}$) subjected to a CEF of cubic symmetry, the eight-fold degeneracy of 4\textit{f}$^{13}$ (\textit{J} = 7/2) for Yb$^{3+}$ and ten-fold degeneracy of 4\textit{f}$^{3}$ (\textit{J} = 9/2) for Nd$^{3+}$ electrons are lifted~\cite{ks6z-6nxj, PhysRevB.110.144434}. The ground state manifold ${}^2F_{7/2}$ for Yb$^{3+}$now splits into two doublets ($\Gamma_6$ and $\Gamma_7$) and a quartet ($\Gamma_8$)~\cite{BIRGENEAU197259,LEA19621381}. The single-ion wave functions for Ba$_2$YbNbO$_6$ are:

$\Gamma_6$ = 0.6455$| \pm\frac{7}{2}\rangle$ + 0.7638$| \pm\frac{1}{2}\rangle$,

$\Gamma_7$ = 0.866$| \pm\frac{5}{2}\rangle$ + 0.5$| \pm\frac{3}{2}\rangle$,

and

$\Gamma_8$ = 0.7638$| \pm\frac{7}{2}\rangle$ + 0.6455$| \pm\frac{1}{2}\rangle$ + 0.5$| \pm\frac{5}{2}\rangle$ + 0.866$| \pm\frac{3}{2}\rangle$.

The doubly degenerated ground state Kramers pairs are dominated by $| \pm\frac{1}{2}\rangle$ of the $\Gamma_6$ wave function with modest mixing of higher  $| \pm m_J\rangle$ as seen earlier, and also indicated by the heat capacity results discussed later. In contrast, the energy eigenvalues and wavefunction for Ba$_2$NdNbO$_6$ are listed in the Table~\ref{Nd_eigen} with a quartet ground state according to the single-ion anisotropy model. The calculated \textit{g}-tensor, originating from the orbital contribution to the ground-state Kramers doublets, is listed in Table~\ref{tensor:g}. However, an octahedral tilting may reduce the local symmetry, resulting in a pseudo-quartet ground state by lifting the four-fold degeneracy. Due to the close proximity of the low-lying crystal-field multiplets, mixing of crystal-field levels can occur, creating a complex crystal-field environment, as observed in NdB$_4$~\cite{PhysRevB.111.104424}. Further exploration is needed to understand the local symmetry and its effect on the crystal field environment.

\begin{table}[t]
\centering
\caption{Ground-state \textit{g}-tensors for (a) Ba$_2$YbNbO$_6$ and (b) Ba$_2$NdNbO$_6$.}
\label{tensor:g}
\begin{tabular}{cc}
\textbf{(a)} & \textbf{(b)} \\[4pt]
$
\begin{pmatrix}
-2.667 & 0 & 0 \\
0 & 2.667 & 0 \\
0 & 0 & 2.667
\end{pmatrix}
$
&
$
\begin{pmatrix}
-3.064 & 0 & 0 \\
0 & -3.064 & 0 \\
0 & 0 & 0.418
\end{pmatrix}
$
\end{tabular}
\end{table}

\subsection{Heat capacity}
\begin{figure*}
      \centering
      \includegraphics[width=0.98\textwidth]{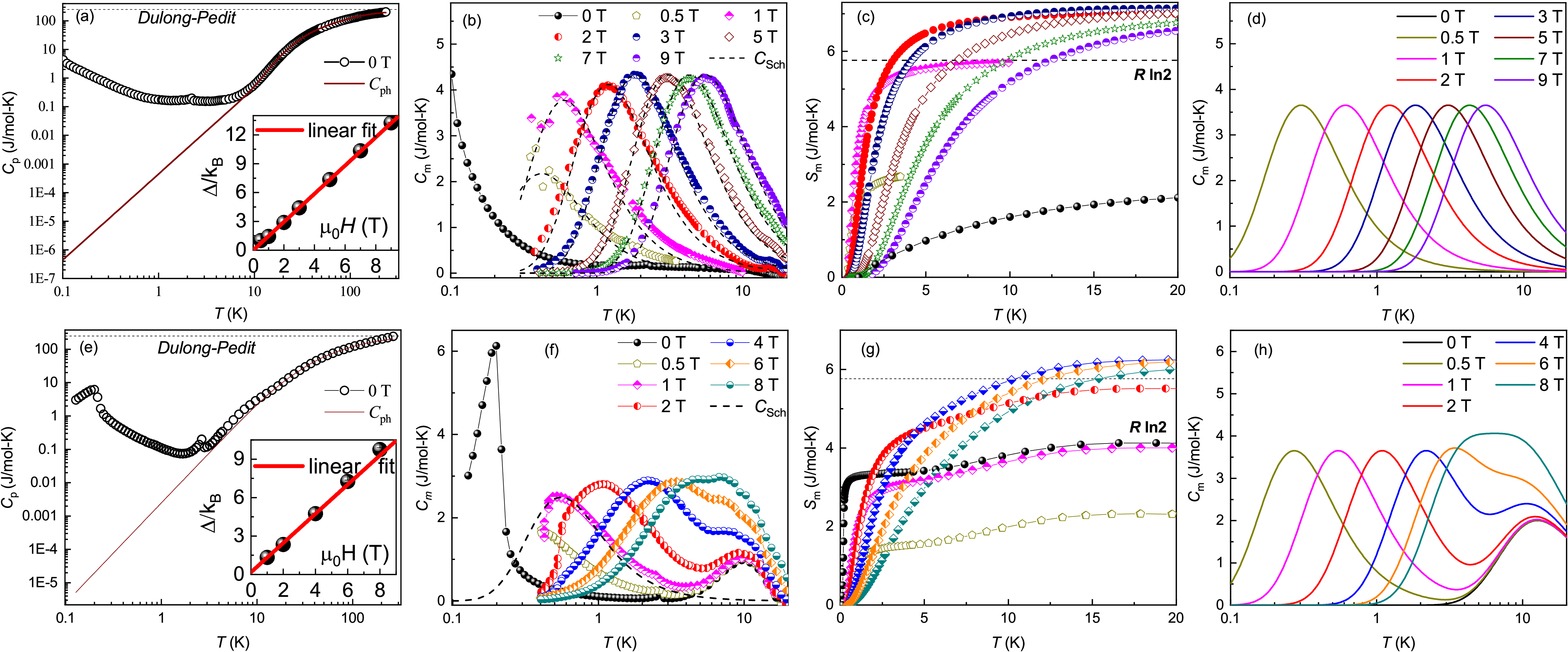}
      \caption{(a) and (e) Zero-field heat capacity ($C_p$) (black open circles) of Ba$_2$YbNbO$_6$ and Ba$_2$NdNbO$_6$, respectively, with lattice contribution ($C_{ph}$) (solid bordeaux line), inset showing the linear-field dependence of the Schottky gap ($\Delta/\mathrm{k}_B$). (b) and (f) Magnetic heat capacity ($C_{mag}$) at zero field and applied fields (0.5~T to 9~T), fitted with Schottky function due to Zeeman splitting of the $j_{\rm eff}$ = 1/2 doublet. $C_{mag}$ for Ba$_2$NdNbO$_6$ reveals a sharp increase at low temperatures, indicative of long-range magnetic ordering (\textit{T}$_N$ = 200~mK) at zero-field. (c) and (g) Magnetic entropy change $\Delta S_{mag}$ obtained by integrating $C_{mag}/T$ vs T, saturating towards \textit{R}ln2 with an applied field of 1~T and higher for Ba$_2$YbNbO$_6$ and Ba$_2$NdNbO$_6$, indicating the $j_{\rm eff}=1/2$ ground state. $\Delta S_{mag} (T)$, showing 61\% of the expected \textit{R}ln2 entropy released down to 0.12~K for Ba$_2$NdNbO$_6$ case. (d) and (h) are the Simulated Schottky behavior using CEF schemes for various applied fields.}
      \label{fig: BYNO_HC}
\end{figure*}

Low temperature heat capacity [$C_{\mathrm p}(T,H)$] was measured down to 0.1~K in zero applied field and down to 0.4~K with different constant applied magnetic fields for Ba$_2$YbNbO$_6$ [Fig.~\ref{fig: BYNO_HC}(a,b)] and Ba$_2$NdNbO$_6$ [Fig.~\ref{fig: BYNO_HC}(e,f)]. The zero-field heat capacity [$C_{\mathrm p}(T)$] data consist of phonon [$C_{\mathrm ph}(T)$] and magnetic [$C_{\mathrm mag}(T)$] contributions. To find the magnetic contributions to the total heat capacity, the lattice contribution is estimated in the temperature region 16~K $\leq T \leq$ 70~K and 14~K $\leq T \leq$ 105~K for Ba$_2$YbNbO$_6$ and Ba$_2$NdNbO$_6$, respectively and subtracted by considering a combination of Debye and Einstein terms~\cite{kittel2004introduction}:
\begin{align}
\label{eq: Debye-Einstein equation}
C_{\rm ph} =f_{\rm D}\left[ 9nR\left(\frac{T}{\theta_{\rm D}} \right)^3\int_{0}^{x_{\rm D}} \frac{x^4e^x}{(e^x - 1)^2}\,dx\right] \nonumber \\+\sum_{i=1}^{3}g_ {\rm E_i}\left[3nR\left(\frac{\theta_{\rm E_i}}{T} \right)^2\frac{exp\left(\frac{\theta_{\rm E_i}}{T} \right)} {\left[exp\left(\frac{\theta_{\rm E_i}}{T} \right)-1\right]^2}\right].
\end{align}
Here, we have considered one Debye and three Einstein terms representing the lattice contribution to the total heat capacity (bordeaux solid line), respectively [Fig.~\ref{fig: BYNO_HC}(a)].

In Eq.~\eqref{eq: Debye-Einstein equation}, $R$ is the universal gas constant, and $n$ (=10) is the number of atoms per formula unit. The Debye temperature ($\theta_{\rm D}$) is estimated to be 162.821~K and 92.179~K for Ba$_2$YbNbO$_6$ and Ba$_2$NdNbO$_6$, respectively, and the contributions of Einstein temperatures ($\theta_{\rm E1}, \theta_{\rm E2}, \theta_{\rm E3}$) are 192.441~K, 435.464~K, and 106.859~K for Ba$_2$YbNbO$_6$, and 155.562~K, 490.723~K, and 240.736 ~K for Ba$_2$NdNbO$_6$. The sum of the adjustable weight factors ($f_{\rm D}$, $g_{\rm E_i}$) is taken to be unity. Due to the presence of low-lying CEF levels, there is always a finite uncertainty in the estimation of phonon contribution as discussed earlier. Therefore, an approximation of the Debye theory at low temperatures $C_{p, lattice}=\beta T^3+\delta T^5$ was implemented and is consistent with the Debye-Einstein model fit. $C_{mag}$ value below 16~K remains the same and is independent of the fitting condition and hence is considered reliable.  


Due to the relatively weak exchange interactions in rare-earth-based compounds, sub-Kelvin measurements are usually necessary to understand the nature of the magnetic ground state. The zero field $C_{\rm mag}(T)$ for Ba$_2$YbNbO$_6$ shows no signature of magnetic LRO down to 0.1~K [Fig.~\ref{fig: BYNO_HC}(a)], resulting in \textit{f}$\approx$~3.3 (frustration parameter, $f=\frac{\theta_\mathrm{CW}}{T_{\rm N}}$) to be the lower limit. However, the sharp divergence of $C_{\rm mag}(T)$ with decreasing temperature indicates the development of magnetic correlations (which match well with the low-temperature magnetization results) and also suggests the presence of magnetic ordering at even lower temperatures. With the magnetic field evolution [Fig.~\ref{fig: BYNO_HC}(b)], the low temperature upturn reduces, and a hump-like feature appears and gradually shifts towards the high temperature region. This indicates typical Schottky behavior, in which an applied magnetic field lifts the degeneracy of the ground-state Kramers doublet. It is important to note that the low-temperature behavior is driven by the pseudospin $\mathrm{\textit{j}_{\rm eff}}=1/2$ due to the CEF, as discussed earlier in the magnetization analysis. Therefore, for further confirmation of $\mathrm{\textit{j}_{eff}}=1/2$ ground state Schottky function Eq.~(\ref{Schottky}) has been implemented as shown by the black dashed line in Fig.~\ref{fig: BYNO_HC}(b).

\begin{equation}
C_{\rm Sch} = \eta R\left[\frac{\Delta(\mu_0H)}{k_{\rm B}T}\right]^2\frac{e^{\left(\frac{\Delta(\mu_0H)}{k_{\rm B} T}\right)}}{\left[1 +  e^{\left(\frac{\Delta(\mu_0H)}{k_B T}\right)}\right]^2}.
\label{Schottky}
\end{equation}
Here, $\eta$ is the fraction of spins involved in the Schottky anomaly, \textit{R} is the universal gas constant, and $\frac{\Delta}{k_B}$ represents the two-level energy gap. $\frac{\Delta}{k_{\rm B}}$ extracted from Schottky fitting, is shown in the inset of Fig.~\ref{fig: BYNO_HC}(a), and the energy gap is linearly increasing with magnetic field as expected for a two-level Schottky system. A linear fit of $\frac{\Delta}{k_{\rm B}}$ passing through the origin rules out the possibility of any internal magnetic field and results in a \textit{g}-value $ \sim 2.3$, consistent with magnetization studies.

By applying a magnetic field ($\mu_0$\textit{H}), Zeeman splitting occurs as $E_{Z}=g\mu_{B}H$ where \textit{g} is the Lande factor, $\mu_B$ is the Bohr magneton, and results in changes in energy levels. Thus, with increasing field strength, the intrinsic antiferromagnetic correlation is suppressed, and Yb$^{3+}$ ions tend to behave as free ions. At high applied field limit (above 1~T in this case), the Schottky fraction ($\eta$) reaches 100~\%, indicating a fully polarized state.

\begin{figure*}
      \centering
      \includegraphics[width=0.98\textwidth]{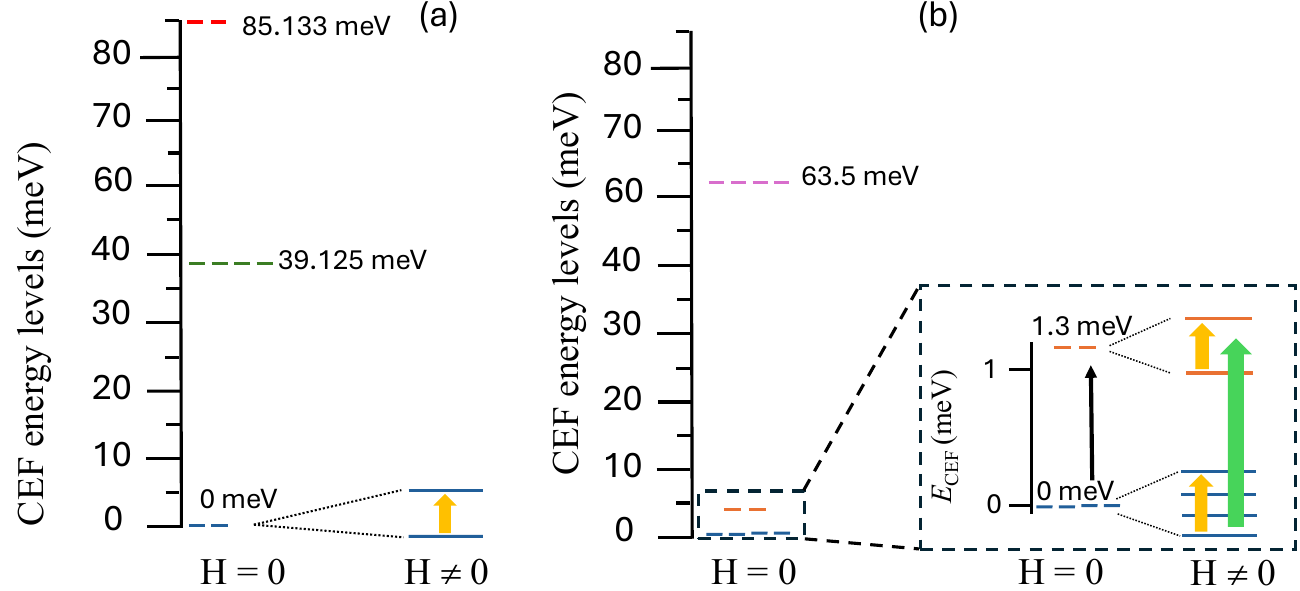}
      \caption{Energy level schemes due to CEF effect at zero applied field (H = 0) for (a) Ba$_2$YbNbO$_6$ (0 meV, 39.12 meV, and 91.88 meV) and (b) Ba$_2$NdNbO$_6$ (0 meV, 1.3 meV, and 63.5 meV) obtained using point charge calculations. In the presence of a magnetic field (H $\neq$ 0), the degeneracy of the lowest Kramers levels is lifted by Zeeman splitting, and the arrows indicate the CEF transitions between the states corresponding to the Schottky peaks in the heat capacity data.}
      \label{fig: CEF_levels}
\end{figure*} 

The temperature dependence of heat capacity for Ba$_2$NdNbO$_6$ is shown in Fig.~\ref{fig: BYNO_HC}(e), in which a decrease in temperature is accompanied by a sharp increment of $C_p(0)$, indicating the magnetic correlation enhances as the compound enters to a long-range magnetic ordered (LRO) state at $T_{\rm N} = 0.198$~K. Therefore, the resultant frustration parameter \textit{f} $\approx$~2.6, indicating an order of frustration similar to Ba$_2$YbNbO$_6$. After subtracting the lattice contribution, the magnetic heat capacity is estimated at zero applied field and also with a varying constant magnetic field up to 9~T as shown in Fig.~\ref{fig: BYNO_HC}(f). An anomaly in the heat capacity around 10~K is observed, arising from the crystal-field contribution. Since the first excited Kramers doublet level is calculated to be 1.3~meV (see Table~\ref{Nd_eigen}) for Ba$_2$NdNbO$_6$, its effect can be expected around 10~K. This further supports the reliability of the single-ion anisotropy calculation of the CEF scheme for Nd$^{3+}$ in Ba$_2$NdNbO$_6$.

To evaluate total magnetic entropy released \textit{S}$_\mathrm{m}$(\textit{T}) associated with $C_{\rm mag}(T)$, an integration of $C_{\rm mag}(T)/T$ over the entire temperature range has been performed for both compounds. For Ba$_2$YbNbO$_6$, as shown in Fig.~\ref{fig: BYNO_HC}(c), 
the zero-field heat capacity data results in an entropy release of 40\% and with an applied field of 1 T, \textit{S}$_\mathrm{m}$(\textit{T}) saturates to 100~\% of \textit{R}ln2, indicating ${j_{\rm eff}} = 1/2$ ground state at low-temperatures as expected. Furthermore, with increasing the magnetic field of 2~T or above, higher CEF levels are populated, thus \textit{S}$_\mathrm{m}$(\textit{T}) deviates from \textit{R}ln2. This further confirms the ground state is governed by the lowest ${j_{\rm eff}} = 1/2$ Kramers doublet. The experimentally observed behavior of $C_\mathrm{mag}(H, T)$ matches very well with the calculated $C_{\rm CEF}$ results shown in Fig.~\ref{fig: BYNO_HC}(d).

For Ba$_2$NdNbO$_6$, the entropy change \textit{S}$_\mathrm{m}$(\textit{T}) from 16~K to 0.12~K is estimated to be 61\% of \textit{R}ln2. To be noted that the heat capacity has been measured down to 0.12~K due to experimental limitations, which is just below the LRO region. Therefore, a finite contribution to the entropy release remains, which will eventually render \textit{S}$_\mathrm{m}$(\textit{T}) close to the expected value of \textit{R}ln2 for $\mathrm{\textit{J}_{\rm eff}}=1/2$ state. Under the field evaluation, the entropy release approaches towards \textit{R}ln2, indicating the quasi-quartet nature of the ground state as discussed earlier.

In all the low-temperature heat capacity data of Ba$_2$NdNbO$_6$ (zero field and field-dependent cases) [Fig.~\ref{fig: BYNO_HC}(g)], a hump-like feature appears around 10~K. In the absence of a magnetic field, the Kramers doublet energy levels are degenerate, and thus the 10~K hump is attributed to the interlevel transition. Now, applying a magnetic field lifts the degeneracy of the Kramers levels. All the expected inter- and intra-level transitions are shown in Fig.~\ref{fig: CEF_levels}(b). Thus, by applying a magnetic field of 0.5~T up to 6~T, a second hump-like feature appears at a lower temperature (below 1~K) and shifted towards the higher temperature side and eventually merges with the high temperature hump for an applied magnetic field of 8~T [Fig.~\ref{fig: BYNO_HC}(f)]. This behavior indicates a Schottky-like effect in which a gap opens in the quartet ground-state and first-excited-state doublets under a magnetic field, and one would expect a linear field dependence of the energy gap. In Fig.~\ref{fig: BYNO_HC}(f), we have fitted (dashed black line) the low temperature hump for 1~T applied field with Eq.~(\ref{Schottky}), and the same procedure has been adapted for the higher field dependence of heat capacity too. The field dependence of the gap ($\Delta/k_B$) is plotted in the inset of Fig.~\ref{fig: BYNO_HC}(e), resembling the expected Schottky effect with \textit{g}-value~1.8, consistent with magnetization studies.

The experimentally observed behavior of $C_\mathrm{mag}(H,T)$ matches very well with the calculated $C_\mathrm{CEF}$ results shown in Fig.~\ref{fig: BYNO_HC}(h). At low field, spectral weight transfer to the low temperature hump is more, and with increasing magnetic field strength, the low-temperature hump is shifted towards the high temperature, resulting in the line broadening of the high-temperature hump, and is consistent with experimental findings~\cite{PhysRevB.110.144434}. Therefore, it is clear that, upon applying a field, the low-temperature feature is due to the transition between the Zeeman levels of the quartet ground state, resulting in Schottky-like behavior. The high-temperature hump is due to the transition from ground states to the first excited state, as shown in Fig.~\ref{fig: BYNO_HC}(f,h). It is evident that the effect of $C_\mathrm{CEF}$ on the total heat capacity from higher CEF levels is not dominant, as the higher excited levels are situated at 63.5~meV (Fig.~\ref{fig: CEF_levels}(b)). Thus, at high temperatures where the heat capacity is mostly dominated by phonon contribution, $C_\mathrm{CEF}$ from the second excited level contribution is small, and higher levels' contributions are beyond the experimental temperature range.


\subsection{First-Principles Study of Magnetism and Electronic Structure}
\begin{figure}[h]
\begin{center}
    \includegraphics[width=1\columnwidth]{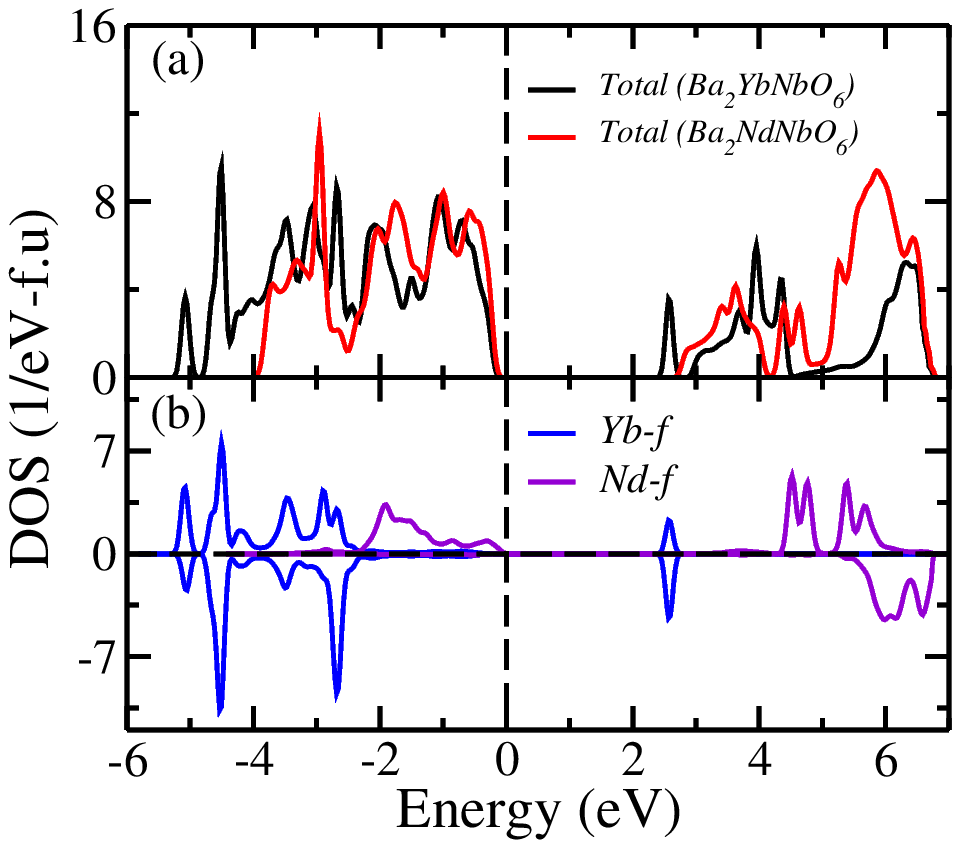}
    \caption{Calculated spin polarized density of states (DOS) including spin-orbit coupling (SOC). (a) A total DOS for Ba$_2$YbNbO$_6$ and Ba$_2$NdNbO$_6$. (b) Spin resolved projected density of states (PDOS) for Yb$-f$, Nd$-f$ orbitals, respectively.}
    \label{dos-nd}
    \end{center}
\end{figure}

We begin by establishing the lowest energy state of Ba$_2$YbNbO$_6$ and Ba$_2$NdNbO$_6$ from total-energy calculations performed within the GGA+$U$ frameworks. For both compounds, the \textcolor{blue} {Type-I} collinear antiferromagnetic (AFM) configuration is found to be marginally lower in energy than the ferromagnetic (FM) state. Importantly, the energy differences between the FM and AFM configurations are extremely small, reflecting the highly localized nature of the rare-earth 4$f$ moments. The Type-I AFM state is found to be the lowest energy  state, with energies lower than the FM configuration by 0.20 meV/f.u. and 0.36 meV/f.u. for Ba$_2$YbNbO$_6$ and Ba$_2$NdNbO$_6$, respectively. In contrast, the Type-II AFM state lies 0.13 meV/f.u. below and 0.17 meV/f.u. above the FM state for Ba$_2$YbNbO$_6$ and Ba$_2$NdNbO$_6$, respectively, while the Type-III AFM state is 0.05 meV/f.u. and 0.16 meV/f.u. higher in energy than the FM state for the two compounds. This near-degeneracy directly reflects the experimentally observed small Curie–Weiss temperatures $\theta_{CW}$, indicating very weak inter-site magnetic exchange interactions and placing both systems in the regime of weakly coupled local moments. Next, we analyzed the electronic structure of Ba$_2$YbNbO$_6$ and Ba$_2$NdNbO$_6$. This is done by examining their total and orbital-projected density of states (DOS) calculated within the GGA+$U$+SOC framework, as shown in Fig.~\ref{dos-nd}. For both compounds, the total DOS (Fig.~\ref{dos-nd}(a)) clearly establishes an insulating ground state, with charge gaps of approximately 2.53~eV for Ba$_2$YbNbO$_6$ and 2.82~eV for Ba$_2$NdNbO$_6$. The presence of a sizable gap, together with the sharp spectral features, is characteristic of systems with strongly localized rare-earth 4$f$ electrons. In both cases, the occupied states below the Fermi level are dominated by rare-earth 4$f$ contributions, while the unoccupied conduction bands are primarily derived from Nb-4$d$ states. Our calculations confirm that Nb remains essentially non-magnetic and plays no direct role in the low-energy magnetic properties of these systems. The orbital-projected DOS shown in Fig.~\ref{dos-nd}(b) highlights the distinct energetic positioning of the rare-earth 4$f$ states in the two compounds. In Ba$_2$YbNbO$_6$, the Yb-4$f$ manifold is predominantly occupied and lies well below the Fermi level, with only a single unoccupied 4$f$ state appearing above it. This is consistent with the 4$f^{13}$ electronic configuration of Yb$^{3+}$ and results in an effective Kramers doublet with ${j_{\rm eff}} = 1/2$, further split by strong spin–orbit coupling. In contrast, for Ba$_2$NdNbO$_6$, the occupied Nd-4$f$ states in the valence band are shifted significantly closer to the Fermi level, while the unoccupied Nd-4$f$ states appear at higher energies in the conduction band. This distribution is consistent with the nominal 4$f^{3}$ configuration of Nd$^{3+}$. The closer proximity of the Nd-4$f$ states to the Fermi level, compared to the deeply localized Yb-4$f$ states, is reflected in the calculated spin and orbital moments and thus plays an important role in shaping the relativistic effects discussed below. Before discussing relativistic effects, we quantified interatomic magnetic interactions ($J_{ij}$) beyond total-energy comparisons. We evaluated the $J_{ij}$ couplings using the magnetic force theorem within the FP-LMTO implementation of the RSPT code~\cite{Rspt2}. The calculated isotropic exchange interaction for Ba$_2$YbNbO$_6$ with the first and second nearest neighbour coupling $J_1=-0.006$meV and $J_2=-0.002$meV, respectively. Whereas for Ba$_2$NdNbO$_6$, $J_1=-0.013$meV and $J_2=-0.004$meV. These values are consistent with the large separation between the rare-earth ions ($>6$~Å) and the highly localized nature of the 4$f$ electrons, which leads to very weak inter-site exchange interactions. The small magnitudes of the exchange couplings are also fully consistent with the very small total-energy differences observed among the competing magnetic configurations in our GGA+$U$+SOC calculations, as discussed. Remarkably, the extracted exchange parameters are found to be negligibly small, within the numerical accuracy of DFT+$U$ for both compounds. This result reinforces the picture of extremely weak magnetic exchange, consistent with the small $\theta_{\mathrm{CW}}$ values as reported experimentally. The spin and orbital moments remain substantial, as summarized in Table~\ref{tab:mae_wide}. For Ba$_2$YbNbO$_6$, the spin and orbital moments along the easy axis are $\mu_S = 0.31~\mu_B$ and $\mu_L = 0.92~\mu_B$, respectively, while Ba$_2$NdNbO$_6$ exhibits significantly larger values of $\mu_S = 3.00~\mu_B$ and $\mu_L = 4.10~\mu_B$, consistent with the CEF findings. The coexistence of negligible inter-site exchange and large on-site orbital moments indicates that crystal electric field (CEF) effects, together with strong spin--orbit coupling (SOC), generate dominant single-ion anisotropy that governs the low-energy physics of these materials. Consistent with this picture, SOC gives rise to a pronounced magnetic anisotropy energy (MAE) in the lowest-energy Type-I AFM state, with the crystallographic (001) direction serving as the easy axis of magnetization (as displayed in Table~\ref{tab:mae_wide}). The calculated MAE is approximately 20~meV for Ba$_2$YbNbO$_6$ and increases substantially to 103~meV for Ba$_2$NdNbO$_6$, in agreement with the estimated $g$-tensor anisotropy shown in Table~\ref{tensor:g}. This behavior arises from the coupling of SOC to the anisotropic 4$f$ charge density within the local crystal-field environment, leading to a direction-dependent magnetic energy even though the overall crystal symmetry is cubic. Similar behavior has been reported in cubic $f$-electron compounds, e.g., Yb$_2$Ti$_2$O$_7$, Tb$_2$Ti$_2$O$_7$, where the interplay of CEF effects and strong SOC produces pronounced magnetic anisotropy despite the preservation of global cubic symmetry~\cite{Tokiwa2016, PhysRevB.62.6496}. Note that the strength of the dipolar interaction, $D = \left(\frac{\mu_{eff}^2}{r^3}\right)$ ($\mu_{eff}$ is effective magnetic moment at paramagnetic state and `$r$' is the distance between the nearest-neighbour ions) for Ba$_2$YbNbO$_6$ and Ba$_2$NdNbO$_6$ compounds are $\approx$35~mK and $\approx$19~mK, respectively, indicating a negligible contribution to the magnetic ground state.  

\begin{table}[t]
\centering
\caption{Calculated magnetocrystalline anisotropy energies and local magnetic moments of rare-earth ions in Ba$_2$YbNbO$_6$ and Ba$_2$NdNbO$_6$. The total energy differences $\Delta E$ are given for spin quantization along the crystallographic (100), (010), and (001) directions. The spin ($\mu_S$) and orbital ($\mu_L$) magnetic moments are reported only for the easy axis of magnetization.}
\vspace{0.1cm}

\setlength{\tabcolsep}{6pt} 
\renewcommand{\arraystretch}{1.6}
\begin{tabular}{l c c c c c c}
\hline \hline
Compound & Ion &
\multicolumn{3}{c}{$\Delta E$ (meV)} &
$\mu_S$ & $\mu_L$ \\
\cline{3-5}
 &  & (100) & (010) & (001) & ($\mu_B$) & ($\mu_B$) \\
\hline
Ba$_2$YbNbO$_6$ & Yb$^{3+}$ & 0     & 0      & $-20$  & 0.31 & 0.92 \\
Ba$_2$NdNbO$_6$ & Nd$^{3+}$ & 0     & 0  & $-103$ & 3.00 & 4.10 \\
\hline \hline
\end{tabular}
\label{tab:mae_wide}
\end{table}

In addition to the isotropic Heisenberg exchange, we computed the full exchange tensor within the GGA+$U$+SOC framework to quantify the strength of anisotropic magnetic interactions. The anisotropic tensor elements, which underpin bond-dependent Kitaev and symmetric off-diagonal interactions, were found to be negligibly small and within the numerical uncertainty of our calculations. Their magnitudes are considerably smaller than the already weak isotropic exchange couplings, indicating that exchange anisotropy plays a minimal role in both compounds. Collectively, these double perovskites provide a compelling platform for exploring SOC-dominated, single-ion–controlled magnetism in localized 4$f$-electron systems.

\section{CONCLUSION}
We have carried out an extensive investigation of the magnetic properties of the double-perovskite compounds Ba$_2$YbNbO$_6$ and Ba$_2$NdNbO$6$, both crystallizing in the \textit{Fm$\bar{3}$m} space group, where the magnetic ions form a geometrically frustrated fcc lattice. The combined effects of strong crystal electric field and spin--orbit coupling in the localized 4$f$ orbitals stabilize a well-isolated $j_{\rm eff}=1/2$ Kramers doublet ground state in both systems, as evidenced by magnetization and heat-capacity measurements, which corroborate the findings from DFT calculations. For Ba$_2$YbNbO$_6$, the sharp upturn in the heat capacity below 300~mK down to our lowest measured temperature of 100~mK indicates the onset of magnetic LRO below 100~mK. Taking 100~mK as the upper limit for the ordering temperature yields a frustration parameter of approximately 3, reflecting a moderate level of frustration. Even Ba$_2$NdNbO$_6$ exhibits moderate frustration with a magnetic ordering at $T_{\rm N} \simeq 200$~mK. The presence of a magnetically ordered ground state is consistent with the theoretical predictions that rare-earth-based fcc lattices with finite Kitaev interactions can host ordered ground states stabilized via the ``order-by-disorder" mechanism~\cite{PhysRevB.95.085132}. Moreover, our GGA+$U$+SOC calculations identify the Type-I AFM state as the lowest energy magnetic state. The calculated anisotropic exchange terms, including the Kitaev and symmetric off-diagonal components, are found to be negligibly small within the numerical accuracy of our calculations. Our results, therefore, demonstrate that Ba$_2$YbNbO$_6$ and Ba$_2$NdNbO$_6$ possess the key ingredients necessary to realize Kitaev-type exchange interactions. This work broadens the landscape of potential Kitaev materials beyond the conventional systems and establishes these 4$f$-based fcc double perovskites as promising model systems. Future neutron-scattering experiments and detailed theoretical modeling will be essential for elucidating the nature of the ordered state, verifying the presence of Kitaev interactions, and exploring the possible emergence of Weyl-like excitations predicted for such systems.

\section{Acknowledgement} 
SMH and MM would like to acknowledge the UGC-DAE Collaborative Research Scheme (ref.~CRS/2021-22/01/393) for funding. SM, RK, and RN would like to acknowledge SERB, India, bearing sanction Grant No.~CRG/2022/000997. S.K.P. acknowledges funding support from a SERB Core Research grant (Grant No.~CRG/2023/003063).


\bibliography{References}
\end{document}